\pdfoutput=1
\documentclass{jpsj3}
\usepackage{txfonts}
\usepackage{cleveref}
\Crefname{equation}{Equation}{Equations}
\crefname{equation}{Eq.}{Eqs.}
\Crefname{figure}{Figure}{Figures}
\crefname{figure}{Fig.}{Figs.}

\usepackage{graphicx}
\usepackage{array}
\usepackage{bm}

\newcolumntype{M}[1]{>{\centering\arraybackslash}m{#1}}

\newcommand{\argmin}{\mathop{\rm argmin}\limits}

\title{Computed Tomography Reconstruction Algorithm Using Markov Random Field Model}
\author{Taiga Shimomiya$^{1}$, Taichi Kusumi$^{1}$, Masayuki Uesugi$^{2}$, Akihisa Takeuchi$^{2}$, Yuki Sada$^{2}$, Hayaru Shouno$^{3}$, and Masato Okada$^{1}$\thanks{Corresponding author. okada@edu.k.u-tokyo.ac.jp}}
\inst{
$^{1}$ Graduate School of Frontier Sciences, The University of Tokyo, 5-1-5 Kashiwanoha, Kashiwa, Chiba 277-8561, Japan\\
$^{2}$ Japan Synchrotron Radiation Research Institute (JASRI), 1-1-1 Kouto, Sayo-cho, Sayo-gun, Hyogo 679-5198, Japan\\
$^{3}$ Graduate School of Informatics and Engineering, The University of Electro-Communications, 1-5-1 Chofugaoka, Chofu, Tokyo 182-8585, Japan
}
\abst{
X-ray computed tomography (CT) reveals the materials' internal structures non-destructively from a tilt series of projected images. Filtered back projection (FBP) is a widely-adopted reconstruction algorithm in CT owing to its small computational cost. Under low-dose or sparse-view conditions, however, FBP often amplifies noise, severely degrading the reconstructed images. In this study, we evaluated the performance of a Bayesian CT reconstruction algorithm based on the Markov random field model under such adverse conditions. Through simulations, we demonstrated that the proposed algorithm shows higher reconstruction performance than FBP under both low-dose and sparse-view conditions. The hyperparameters are estimated by minimizing the Bayesian free energy, enabling adaptive reconstruction that reflects the noise characteristics of the observed projected image. These results suggest that the proposed algorithm can broaden the applicability of CT to dose-sensitive applications and time-constrained measurements, where only limited observed projected images are available.
}

\begin{document}
\maketitle

\section{Introduction}
X-ray computed tomography (CT) is an imaging technique that visualizes the internal structures of materials non-destructively from a tilt series of projected images. Owing to its non-destructive nature, X-ray CT is widely used for various applications, including medical diagnosis, materials characterization, and industrial inspection \cite{Kak2001, Herman2009}. 
In practical applications, however, the number of projections or the exposure time is often limited to reduce radiation damage, particularly for beam-sensitive materials. Under such low-dose or sparse-view conditions, image reconstruction becomes challenging \cite{Natterer2001, Kudo2013}.
For image reconstruction in X-ray CT, filtered back projection (FBP), which reconstructs an image by filtering in the frequency domain followed by back projection, has been widely adopted due to its low computational cost and straightforward implementation \cite{Kak2001, Natterer2001, Willemink2019} (for details, see Appendix). 
However, since FBP enhances high-frequency components during reconstruction, it amplifies noise in the observed projected image, leading to the severe degradation of reconstruction performance under these constrained conditions \cite{Herman2009, Natterer2001}. 
Although various frequency filters, such as Ramp\cite{Ramachandran1971}, Shepp--Logan\cite{Shepp1974}, and Hamming\cite{Harris1978} filters, have been proposed to mitigate this issue, their performance strongly depends on experimental conditions and noise characteristics, making it difficult to guarantee reconstruction performance theoretically.

Under low-dose or sparse-view conditions, it is insufficient to rely solely on the observed projected image, and the prior knowledge about the underlying structures should be incorporated for high-performance reconstruction. 
One effective approach is to utilize Bayesian framework, which allows us to introduce an image-generation model as a prior distribution that captures the underlying structures and to integrate it with the observation process\cite{Lange1995,Geman1984}.
Among various prior distributions, the Markov random field (MRF) prior, which captures the underlying smooth structure by introducing the spatial correlation between the neighboring pixels, has been widely used for image reconstruction\cite{Li2009}.
Furthermore, when the observation noise is assumed to follow the Gaussian distribution and the MRF prior is introduced, the posterior distribution and the free energy can be evaluated analytically, which enables fast image reconstruction.
Within this framework, a series of studies by Nakanishi and others established a method for hyperparameter estimation based on free-energy analysis \cite{Nakanishi2014, Sakamoto2016, Katakami2017}.
Recently, Kusumi \textit{et al}. applied the Bayesian inference based on the MRF model to scanning transmission electron microscopy (STEM) image reconstruction, analytically derived the reconstructed images, and demonstrated its effectiveness through real-time reconstruction \cite{Kusumi2023, Kusumi2024}.

In X-ray CT, Shouno and Okada proposed the Bayesian reconstruction algorithm based on the MRF model, and they demonstrated its high robustness against noise, which had been a major limitation of the FBP method \cite{Shouno2010}.
However, previous studies on CT image reconstruction have mainly focused on the effects of added noise intensity, while comparisons with frequency filters used in FBP and evaluations under reduced projection conditions remain limited.
In this study, we integrate the Bayesian CT reconstruction framework of Shouno and Okada with the hyperparameter estimation method based on the free-energy analysis developed by Nakanishi \textit{et al}., and extend to CT reconstruction the analytical estimation framework demonstrated by Kusumi \textit{et al}.
This enables a CT image reconstruction method that maintains high reconstruction performance not only against noise but also under sparse-projection conditions.
Furthermore, through qualitative and quantitative comparisons with FBP using Ramp, Shepp--Logan, and Hamming filters, we demonstrate that the proposed algorithm achieves superior reconstruction performance over a range of noise intensities and projection numbers.

\section{Theory}

In this section, we formulate a CT reconstruction algorithm based on the MRF model.
The formulation follows the previous study on Bayesian CT reconstruction, where the Gaussian noise is adopted as the noise model and the MRF model is introduced as the prior distribution \cite{Shouno2010}.
Here, we denote the ground truth by $\xi(x,y)$ and the reconstructed image by $f(x,y)$, respectively.

\begin{figure}[htbp]
  \centering
  \includegraphics[width=0.55\columnwidth]{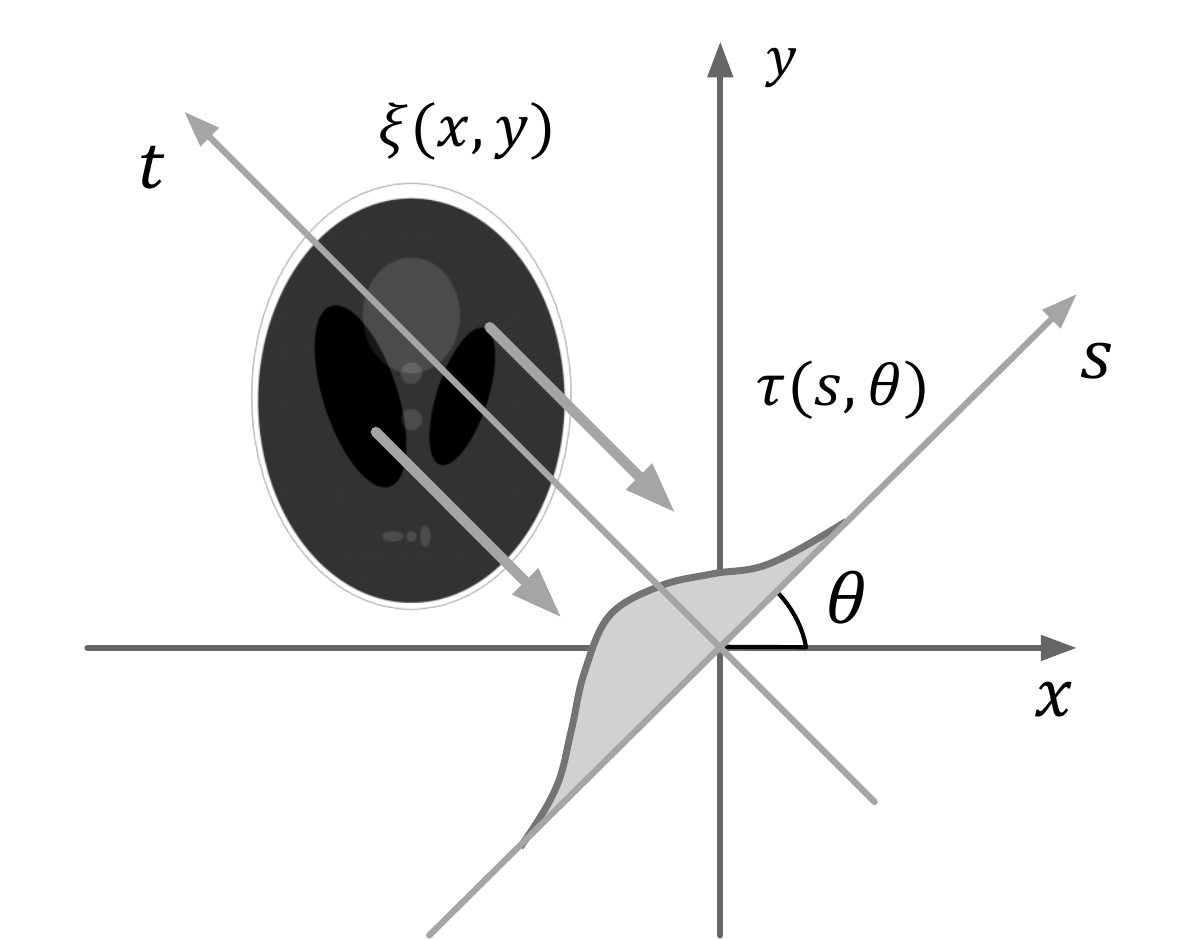}
  \caption{Schematic diagram of the Radon transform used in this study. The object  $\xi(x,y)$ is integrated along lines parallel to $t$ to obtain the projection $\tau(s,\theta)$ on the detector axis $s$, which forms an angle $\theta$ with the $x$ axis.}
  \label{theory}
\end{figure}

The geometry adopted in this study is illustrated in \cref{theory}. For each view angle $\theta$, the detector coordinate $s$ is defined along an axis rotated by $\theta$ with respect to the fixed $(x,y)$ coordinates, and $t$ denotes the coordinate along the direction orthogonal to $s$. The fixed coordinates $(x,y)$ are related to the rotated coordinates $(s,t)$ as
\begin{equation}\label{coord}
  \begin{aligned}
    x &= s\cos\theta - t\sin\theta,\\
    y &= s\sin\theta + t\cos\theta .
    \end{aligned}
\end{equation}
The projection $\tau(s,\theta)$ is obtained by integrating $\xi(x,y)$ along the direction $t$ at fixed $s$ and $\theta$.

For each $\theta$, the noiseless Radon transform \cite{Radon1986} is defined as the line integral of $\xi(x,y)$ along the direction $t$ at fixed $s$.
Assuming additive white Gaussian noise, the observed projected image $\tau(s,\theta)$ is modeled as
\begin{equation}\label{obs}
    \tau(s,\theta) = \int dt\,\xi(s\cos\theta - t\sin\theta,\, s\sin\theta + t\cos\theta) + N(s,\theta),
\end{equation}
where $N(s,\theta)$ follows a Gaussian distribution with the mean of 0 and the precision of $\gamma$.
Accordingly, the likelihood of the observed projected image in Fourier space is given by
\begin{equation}
    p(\boldsymbol{\tau} \mid \boldsymbol{f}, \gamma) \propto \exp\left( -4\pi^2 \gamma \int d\theta \int d\tilde{s} \left| \tilde{\tau}_{\tilde{s}, \theta} - \tilde{f}_{\tilde{s}, \theta} \right|^2 \right).
\end{equation}
Here, $\tilde{\tau}_{\tilde{s}, \theta}$ and $\tilde{f}_{\tilde{s}, \theta}$ denote the Fourier-space representations of $\tau(s,\theta)$ and $f(x,y)$, respectively.
In this study, we introduce a prior distribution which incorporates the assumption  that the ground truth is smooth and does not contain extreme values.
The prior distribution is given by
\begin{equation}\label{prior}
    p(\bm{f}) \propto \exp\left[ - \beta \iint dx\,dy\, \left| \nabla f(x,y) \right|^2 - 4\pi^2 h \iint dx\,dy\, \left| f(x,y) \right|^2 \right].
\end{equation}
The first term in the exponential function imposes the smoothness between neighboring pixels, the second term corresponds to amplitude suppression (L2 regularization), and $\beta$ and $h$ are hyperparameters.
In the polar representation in the Fourier space, the prior distribution is given by
\begin{equation}
    p(\bm{f} \mid \beta, h) \propto \exp\left(-4\pi^2 \int d\theta \int d\tilde{s}\, \left( \beta \tilde{s}^2 + h \right) |\tilde{s}| \left| \tilde{f}_{\tilde{s}, \theta} \right|^2\right).
\end{equation}
By Bayes' theorem, the posterior distribution of the ground truth is given by
\begin{equation}
    p(\bm{f} \mid \bm{\tau}, \gamma, \beta, h) = \frac{p(\bm{\tau} \mid \bm{f}, \gamma)p(\bm{f} \mid \beta, h)}{p(\bm{\tau} \mid \gamma, \beta, h)}.
\end{equation}
Thus, the posterior distribution in the Fourier space is given by
\begin{equation}\label{post}
    \begin{split}
        p(\boldsymbol{f} \mid \boldsymbol{\tau}, \gamma, \beta, h) 
            & \propto \exp\left[ -4\pi^2 \int d\theta \int d\tilde{s} \left( \gamma \left| \tilde{\tau}_{\tilde{s}, \theta} - \tilde{f}_{\tilde{s}, \theta} \right|^2 + (\beta \tilde{s}^2 + h) |\tilde{s}| \left| \tilde{f}_{\tilde{s}, \theta} \right|^2 \right) \right]\\
            &\propto \exp\left[ -4\pi^2 \int d\theta \int d\tilde{s} \left( F_{\tilde{s}} \left| \tilde{f}_{\tilde{s}, \theta} -  \frac{\gamma}{F_{\tilde{s}}}\tilde{\tau}_{\tilde{s}, \theta} \right|^2 + \gamma \left( 1 - \frac{\gamma}{F_{\tilde{s}}} \right) \left| \tilde{\tau}_{\tilde{s}, \theta} \right|^2 \right) \right].
    \end{split}
\end{equation}
where
\begin{equation}\label{fs}
    F_{\tilde{s}} = (\beta \tilde{s}^2 + h)|\tilde{s}| + \gamma.
\end{equation}
We reconstruct the ground truth by maximum a posteriori (MAP) estimation:
\begin{equation}
    \hat{\bm{f}}
    = \arg\max_{\bm{f}} p(\bm{f} \mid \bm{\tau}, \gamma, \beta, h).
\end{equation}
This optimization can be solved analytically, and the optimal reconstructed image in Fourier space is given by
\begin{equation}\label{map}
    \tilde{f}_{\tilde{s}, \theta} = \frac{\gamma}{F_{\tilde{s}}} \tilde{\tau}_{\tilde{s}, \theta}.
\end{equation}

\Cref{map} shows that the proposed algorithm can be interpreted as frequency-domain filtering of the observed projection data, where the factor $\gamma/F_{\tilde{s}}$ plays the role of a reconstruction filter.
As shown in \cref{fs}, since $F_{\tilde{s}}$  $(=(\beta \tilde{s}^2+h)|\tilde{s}|+\gamma)$ depends on the hyperparameters $\beta,h$, and $\gamma$, the filter shape is adaptively determined from the observed projected image through hyperparameter estimation.
To compute \cref{map}, however, the hyperparameters $\gamma$, $\beta$, and $h$ need to be tuned in advance.
In this study, we optimize these hyperparameters by maximizing the marginal likelihood (maximum likelihood estimation) \cite{Bishop2006}.
By the definition of conditional probability, the likelihood of the hyperparameters $p(\bm{\tau} \mid \gamma,\beta,h)$ is given by
\begin{equation}\label{a22}
    \begin{aligned}
        p(\bm{\tau} \mid  \gamma,\beta, h) &= \int d\bm{f}\, p(\bm{f},\bm{\tau} \mid  \gamma,\beta, h) \\
                      &= \int d\bm{f}\, \frac{p(\bm{f},\bm{\tau},\gamma,\beta, h)}{p(\gamma,\beta, h)} \\
                      &= \int d\bm{f}\, p(\bm{\tau} \mid  \bm{f},\gamma)p(\bm{f} \mid  \beta, h).
    \end{aligned}
\end{equation}
Since the above formulation is defined in continuous Fourier space, we discretize the integrals into finite sums over sampled frequency points to enable analytical evaluation and numerical implementation.
Specifically, the Fourier space is discretized into $N_\theta$ points in the angular direction and $N_{\tilde{s}}$ points in the radial-frequency direction, with sampling intervals $\Delta_\theta$ and $\Delta_{\tilde{s}}$, respectively.
Therefore, $p(\tilde{\bm{\tau}} \mid \gamma, \beta, h)$ is expressed as follows:
\begin{equation}\label{a23}
    \begin{aligned}
        p(\tilde{\bm{\tau}} \mid  \gamma, \beta,h) &\propto \int d\tilde{\bm{f}}
        \exp\left[
            -4\pi^2 \frac{\Delta_\theta \Delta_{s}}{N_s} \sum_{l=0}^{N_{\theta} - 1} \sum_{\tilde{k}=0}^{N_{\tilde{s}}-1} 
            \left( 
               F_{\tilde{s}} \left| \tilde{f}_{\tilde{k},l} -  \frac{\gamma}{F_{\tilde{s}}}\tilde{\tau}_{\tilde{k},l} \right|^2  +  \gamma \left( 1 - \frac{\gamma}{F_{\tilde{s}}} \right) \left| \tilde{\tau}_{\tilde{k},l} \right|^2
            \right) 
       \right]\\
                            &= \prod_{\tilde{k},l} \sqrt{\frac{N_s}{4\pi\Delta_\theta \Delta_s F_{\tilde{s}}}} \exp \left(-4\pi^2 \frac{\Delta_\theta \Delta_{s}}{N_s}\gamma \left( 1 - \frac{\gamma}{F_{\tilde{s}}} \right) \left| \tilde{\tau}_{\tilde{k},l} \right|^2 \right) \\
                            &\propto \prod_{\tilde{k},l} \sqrt{\frac{8\pi \Delta_\theta \Delta_{s}}{N_s}\gamma \left( 1 - \frac{\gamma}{F_{\tilde{s}}} \right)} \exp \left(-4\pi^2 \frac{\Delta_\theta \Delta_{s}}{N_s}\gamma \left( 1 - \frac{\gamma}{F_{\tilde{s}}} \right) \left| \tilde{\tau}_{\tilde{k},l} \right|^2 \right).
    \end{aligned}
\end{equation}
For numerical stability, we define the negative log-likelihood, i.e., the Bayesian free energy, $F(\gamma, \beta, h  \mid \bm{\tilde{\tau}})$, and optimize the hyperparameters by minimizing it:
\begin{equation}
    \tilde{\gamma}, \tilde{\beta}, \tilde{h} = \argmin_{\gamma, \beta, h} F(\gamma, \beta, h  \mid \bm{\tilde{\tau}}),
\end{equation}
where
\begin{equation}
    \begin{split}
        F(\gamma, \beta, h  \mid \bm{\tilde{\tau}}) 
        & \equiv - \log p(\tilde{\bm{\tau}} \mid \gamma, \beta,h) \\
        & = -\frac{1}{2}\sum_{\tilde{k},l} \left( \log \frac{8\pi \Delta_\theta \Delta_{s}}{N_s}\gamma \left( 1 - \frac{\gamma}{F_{\tilde{s}}} \right) - \frac{8\pi^2 \Delta_\theta \Delta_{s}}{N_s}\gamma \left( 1 - \frac{\gamma}{F_{\tilde{s}}} \right)|\tilde{\tau}_{\tilde{k},l}|^2 \right) + \text{const}.
    \end{split}
\end{equation}

\section{Results}
\begin{figure}[htbp]
  \centering
  \includegraphics[width=0.6\columnwidth]{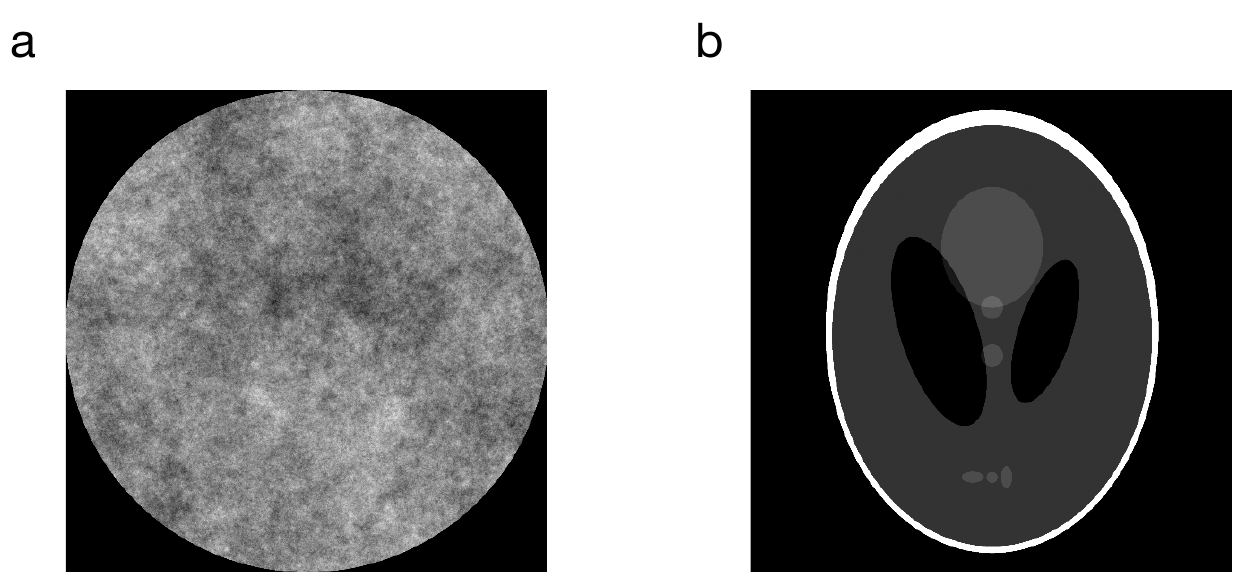}
  \caption{Ground-truth images used in the numerical experiments. (a) Image generated from the MRF model corresponding to the first term of \cref{prior}. (b) Shepp--Logan phantom \cite{Shepp1974, Kak2001}.}
  \label{true}
\end{figure}

\begin{figure}[htbp]
  \centering
  \includegraphics[width=0.95\columnwidth]{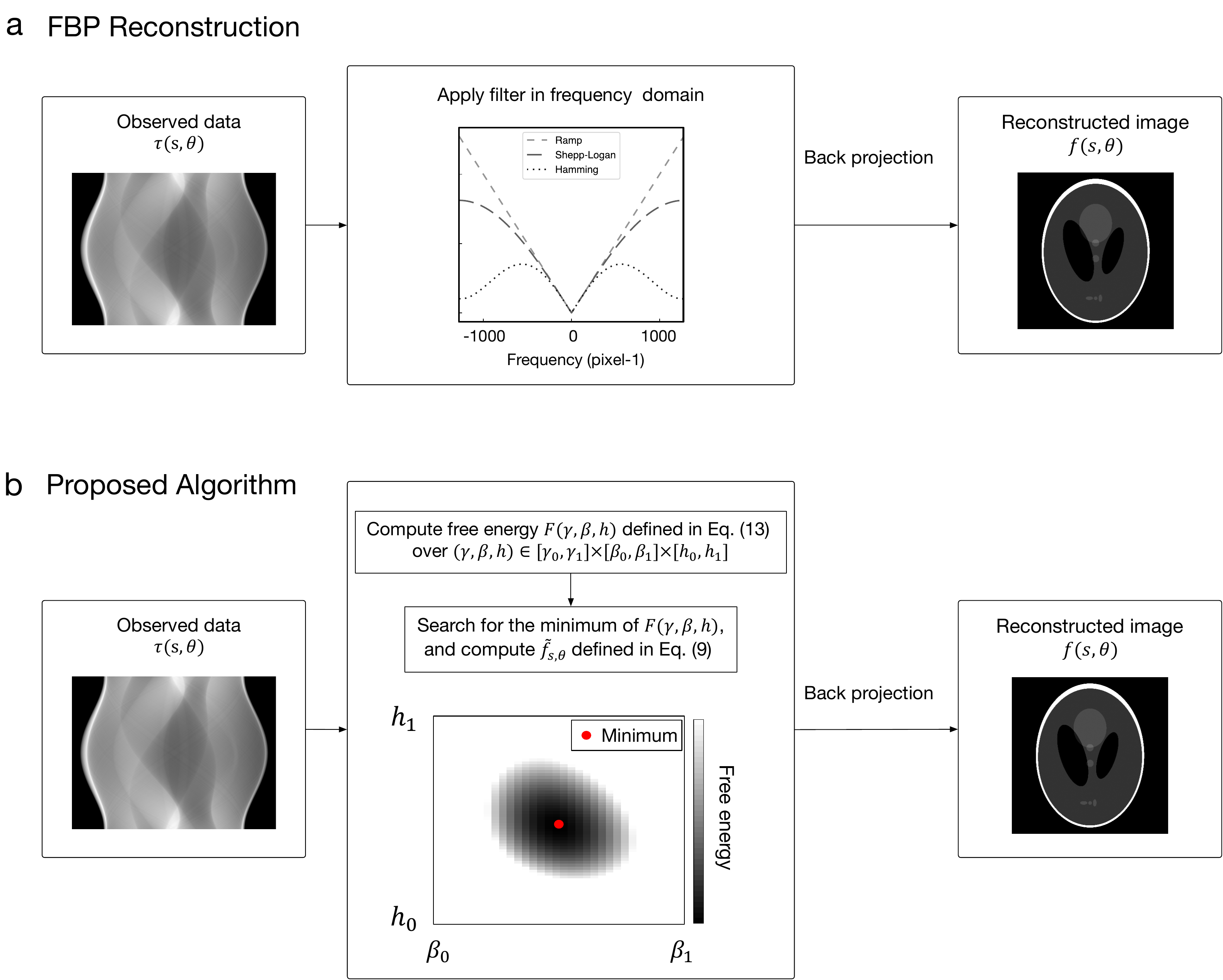}
  \caption{(Color online) Reconstruction procedures in the numerical experiments. (a) FBP: from the observed projected image $\tau(s,\theta)$, frequency-domain filters (Ramp, Shepp--Logan, and Hamming) are applied, and the image is reconstructed by back projection. (b) Proposed algorithm: the hyperparameters $(\gamma, \beta, h)$ are estimated by minimizing the Bayesian free energy $F(\gamma, \beta, h \mid \bm{\tilde{\tau}})$ via grid search; the MAP estimate in \cref{map} is then evaluated in the frequency domain, followed by back projection.}
  \label{flow}
\end{figure}

To verify the effectiveness of the proposed algorithm, we compared its reconstruction performance with FBP reconstruction.
In the numerical experiments, we used two types of ground-truth images: an image generated from the MRF model corresponding to the first term of \cref{prior} and a Shepp--Logan phantom image \cite{Shepp1974, Kak2001}.
We selected these two images to evaluate the proposed algorithm under both model-matched and more general conditions.
The MRF-based image serves as a reference case in which the ground-truth image is statistically consistent with the assumed prior, whereas the Shepp--Logan phantom serves as a more general ground-truth image containing sharp edges.
By varying the noise intensity added to the observed projected images and the number of projections, we evaluated the noise robustness of each algorithm and its stability under limited-projection conditions.
\Cref{true} shows the ground-truth images used in the experiments, and \cref{flow} illustrates the reconstruction workflows for FBP and the proposed algorithm.

In all experiments, the image size and acquisition setting were fixed as follows: $N_x=N_y=N_s=2048$, and the projection angles were uniformly sampled over $180^\circ$ with $N_\theta=1800$ for the full-projection setting, corresponding to an angular interval of $0.1^\circ$.
The hyperparameters $(\gamma, \beta, h)$ in the proposed algorithm were estimated by a grid search of the free energy $F(\gamma, \beta, h \mid \tilde{\bm{\tau}})$.
Specifically, the search region in the $(\gamma, \beta, h)$ space was first divided into a coarse grid, the free energy was evaluated at each grid point, and the neighborhood yielding the minimum was further subdivided into a finer grid.
By repeating this coarse-to-fine search, the optimal hyperparameters were determined efficiently \cite{Kusumi2023}.
In the proposed algorithm, the MAP estimate [\cref{map}] is obtained analytically in Fourier space, and the reconstructed image is computed by frequency-domain filtering of the observed projected image followed by back projection.
This computational structure is equivalent to that of FBP, and because the free energy can also be evaluated analytically, the additional cost for hyperparameter estimation is limited to the number of free-energy evaluations required in the grid search.
For comparison, we used three frequency filters in FBP: Ramp, Shepp--Logan, and Hamming.

In this study, we focused on evaluating the noise-suppression effect of the proposed algorithm and set the experimental conditions as follows.
Changes in the noise intensity correspond to low-dose conditions, whereas a reduction in the number of projections corresponds to sparse-projection conditions.
In particular, in the experiments on projection reduction, we aimed to evaluate how effectively the proposed algorithm suppresses noise when the number of projections is reduced under noisy conditions caused by low dose.
Specifically, for the evaluation of noise intensity, we considered the range $\sigma_{\mathrm{noise}} = 1.0,\,2.0,\,4.0$.
For the projection-reduction experiments, we fixed $\sigma_{\mathrm{noise}} = 2.0$, for which the influence of noise is pronounced in FBP, and gradually reduced the number of projections as $N_\theta = 1800,\,900,\,450$.
When the number of projections is reduced substantially, streak artifacts caused by insufficient angular information are expected to appear.
Because such streak artifacts arise from a mechanism different from noise amplification during reconstruction, in this study we focused on a range of conditions in which the noise-suppression effect can be evaluated clearly.

\subsection{Reconstruction of the synthetic image generated by MRF model}
\subsubsection{Reconstruction performance under various noise intensity}
\begin{figure}[htbp]
  \centering
  \includegraphics[width=0.9\columnwidth]{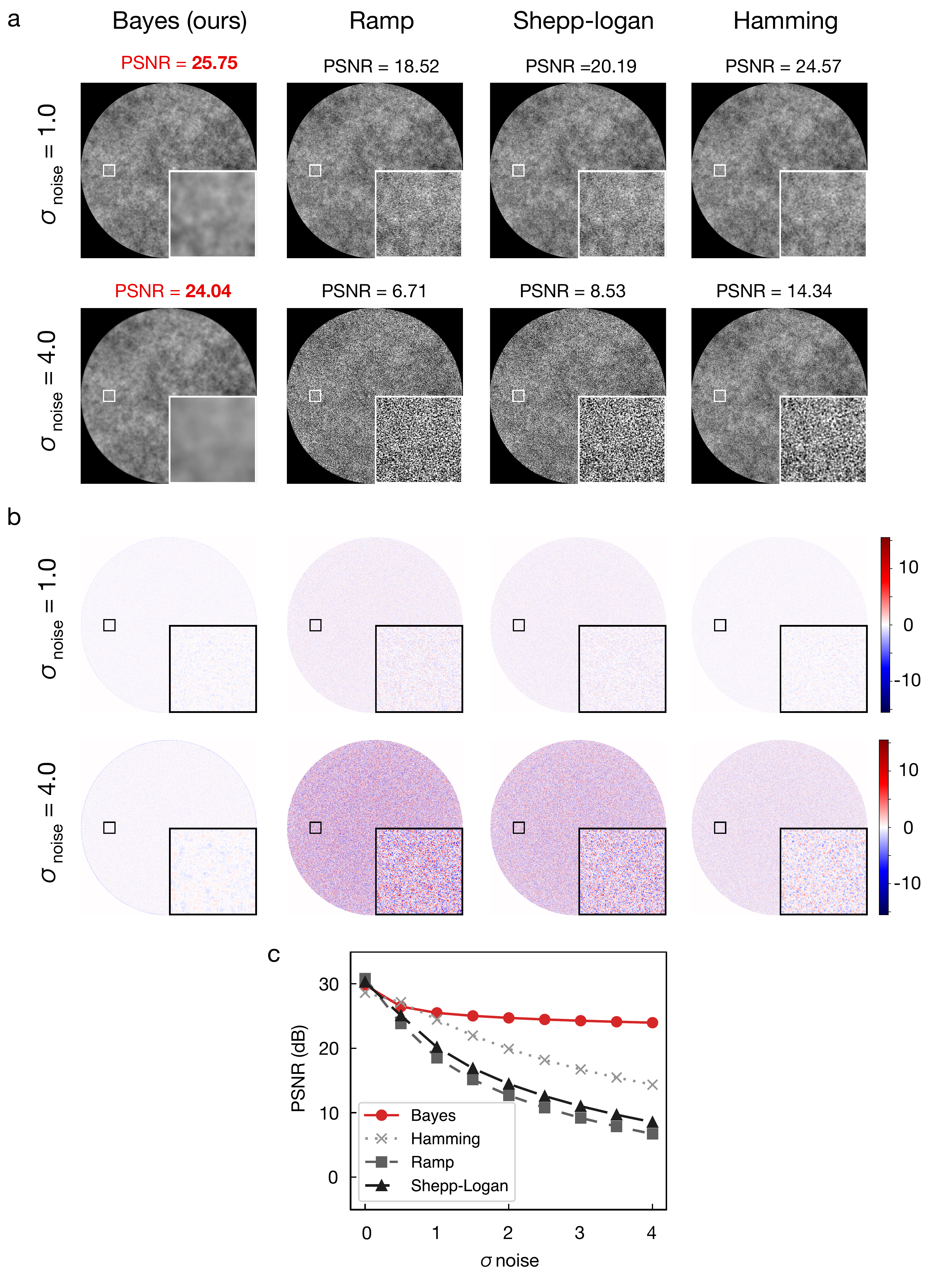}
  \caption{(Color online) Reconstruction performance obtained by fixing the number of projections at $N_\theta = 1800$ and varying the noise intensity. 
  (a) Reconstruction results of the proposed algorithm and FBP for the representative noise conditions $\sigma_{\mathrm{noise}} = 1.0$ and $4.0$, and magnified views of the regions indicated by the boxes. 
  (b) Error images, defined as the reconstructed image minus the ground truth. The color bars indicate the signed error scale. 
  (c) PSNR versus noise intensity for the each method.}
  \label{mrf_noise}
\end{figure}

We first evaluate the reconstruction performance by varying the noise intensity with a fixed number of projections ($N_\theta = 1800$).
\Cref{mrf_noise}(a) shows the reconstructed images by the FBP and the proposed algorithm, together with magnified views of the regions indicated by the boxes. 
The full reconstructed images show that every FBP reconstruction degrades as $\sigma_{\mathrm{noise}}$ increases.
In particular, at the high-noise condition ($\sigma_{\mathrm{noise}} = 4.0$), granular noise appears throughout the field in each FBP reconstruction, as clearly observed in the magnified views.
\Cref{mrf_noise}(b) shows error images, defined as the reconstructed image minus the ground truth, for each noise intensity, together with magnified views of the regions indicated by the boxes.
Since the MRF-based image contains no pronounced edge structure, these error images directly visualize the noise-induced fluctuations in each reconstruction.
As $\sigma_{\mathrm{noise}}$ increases, the FBP reconstructions exhibit increasingly pronounced spatial error fluctuations.
In particular, at $\sigma_{\mathrm{noise}} = 4.0$, large granular errors spread throughout the field for all three FBP filters.
The magnified views show that these fluctuations are amplified in the FBP reconstructions as $\sigma_{\mathrm{noise}}$ increases, whereas the proposed algorithm maintains smaller error amplitudes.
These observations demonstrate the robustness of the proposed algorithm against noise.

To quantify the reconstruction performance, we use the peak signal-to-noise ratio (PSNR).
Let $\mathrm{MSE}$ denote the mean squared error between the reconstructed image and the ground truth.
PSNR is defined as
\begin{equation}\label{psnr}
    \mathrm{PSNR} = 10 \log_{10}\frac{I_{\max}^2}{\mathrm{MSE}},
\end{equation}
where $I_{\max}$ is the maximum possible pixel value.
For each $\sigma_{\mathrm{noise}}$, the proposed algorithm achieves higher PSNR than any of the FBP-based reconstructions compared in \cref{mrf_noise}(a).
Moreover, the rate of PSNR degradation as the noise level increases differs markedly across the methods.
\Cref{mrf_noise}(c) shows the FBP-based filters exhibit a pronounced decrease in PSNR as the noise level increases.
Although the Hamming filter mitigates PSNR degradation to some extent in the high-noise regime, the improvement remains limited compared with the proposed algorithm.
Thus, the proposed algorithm maintains high reconstruction performance over the entire range of noise intensities examined here.

\subsubsection{Reconstruction performance under limited number of projections}
\begin{figure}[htbp]
  \centering
  \includegraphics[width=0.85\columnwidth]{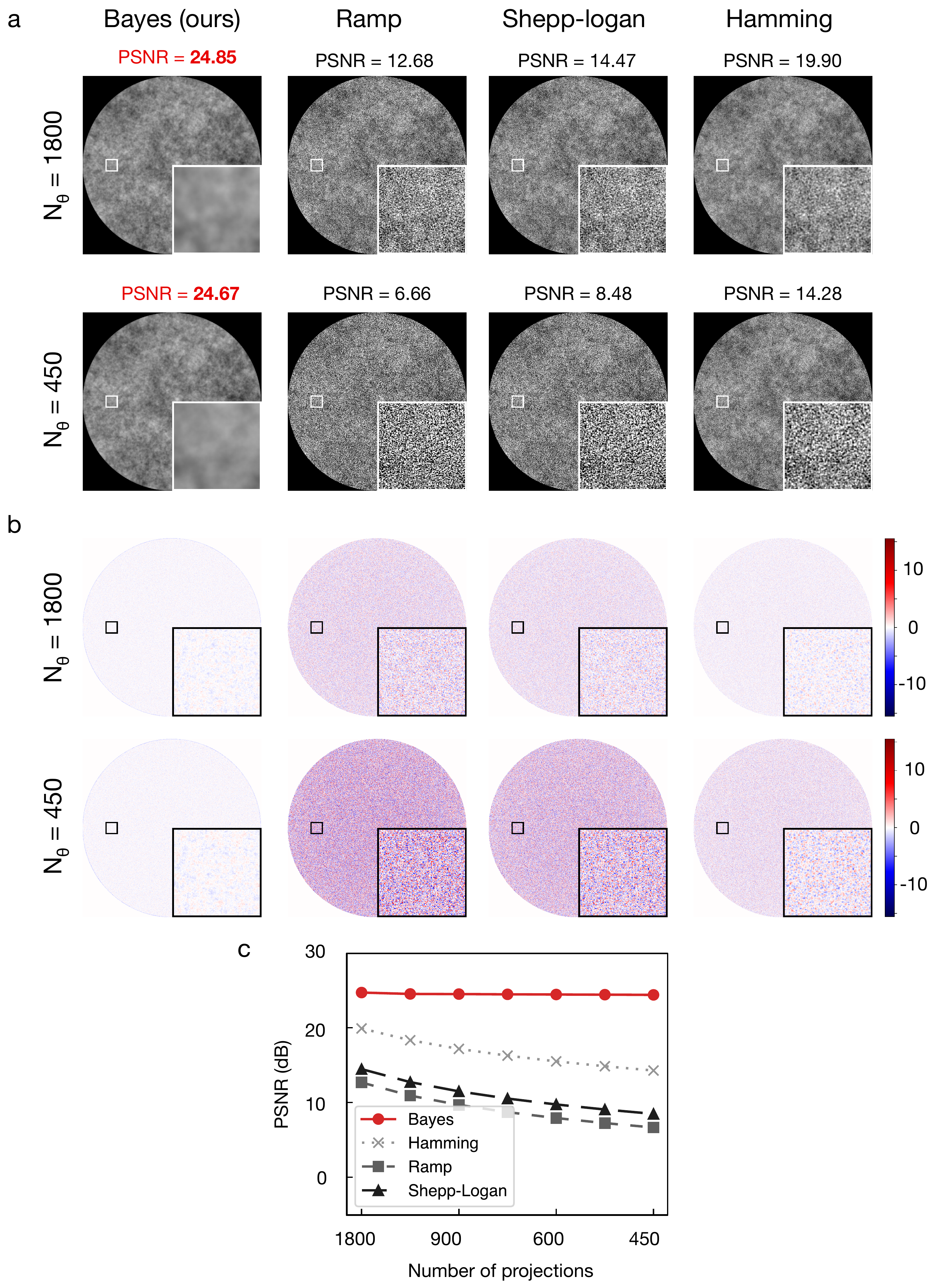}
  \caption{
    (Color online) Reconstruction performance obtained by fixing the noise intensity and varying the number of projections. 
    (a) Reconstruction results of the proposed algorithm and FBP for the representative conditions $N_\theta = 1800$ and $450$, and magnified views of the regions indicated by the boxes. 
    (b) Error images, defined as the reconstructed image minus the ground truth. The color bars indicate the signed error scale. 
    (c) PSNR versus the number of projections for the each method.
    }
  \label{mrf_projection}
\end{figure}

Next, we evaluate the reconstruction performance by reducing the number of projections with the fixed noise intensity ($\sigma_{\mathrm{noise}} = 2.0$).
\Cref{mrf_projection}(a) shows the reconstructed images and the magnified views of the regions indicated by the boxes.
The full reconstructed images show that the reconstruction performance of the FBP-based reconstructions degrades as the number of projections decreases.
Although Gaussian noise with identical variance is added to each projection, reducing the number of projections lowers the angular sampling density, making the reconstruction more sensitive to noise.
The magnified views show that fluctuations become more pronounced in the FBP reconstructions as $N_\theta$ decreases, whereas the proposed algorithm remains more stable.

\Cref{mrf_projection}(b) shows error images, defined as the reconstructed image minus the ground truth, for each projection condition, together with magnified views of the regions indicated by the boxes.
Because the MRF-based image contains no pronounced edge structure, these error images directly visualize the fluctuations caused by reduced angular sampling.
As $N_\theta$ decreases, the FBP reconstructions exhibit increasingly pronounced error fluctuations over the entire field.
The magnified views show that these fluctuations are amplified in the FBP reconstructions, whereas the proposed algorithm maintains smaller error amplitudes.
These observations demonstrate that the proposed algorithm mitigates the degradation caused by reduced angular sampling more effectively than FBP.

To quantify the effect of projection reduction on reconstruction performance, we compare the PSNR values for all methods.
Comparing the PSNR values in \cref{mrf_projection}(a), the proposed algorithm yields higher PSNR than all FBP-based filters for each projection-number condition.
Moreover, the extent of PSNR degradation with decreasing $N_\theta$ differs markedly between the methods.
For all FBP-based filters, PSNR decreases substantially when $N_\theta$ is reduced from $1800$ to $450$.
By contrast, the proposed algorithm exhibits only a small decrease in PSNR and maintains high reconstruction performance.
\Cref{mrf_projection}(c) shows that the proposed algorithm undergoes only a gradual decrease in PSNR as the number of projections is reduced.
These results indicate that the proposed algorithm is more robust to projection reduction than FBP.

\subsection{Reconstruction of Shepp--Logan phantom}
\subsubsection{Reconstruction performance under various noise intensity}
\begin{figure}[htbp]
  \centering
  \includegraphics[width=0.9\columnwidth]{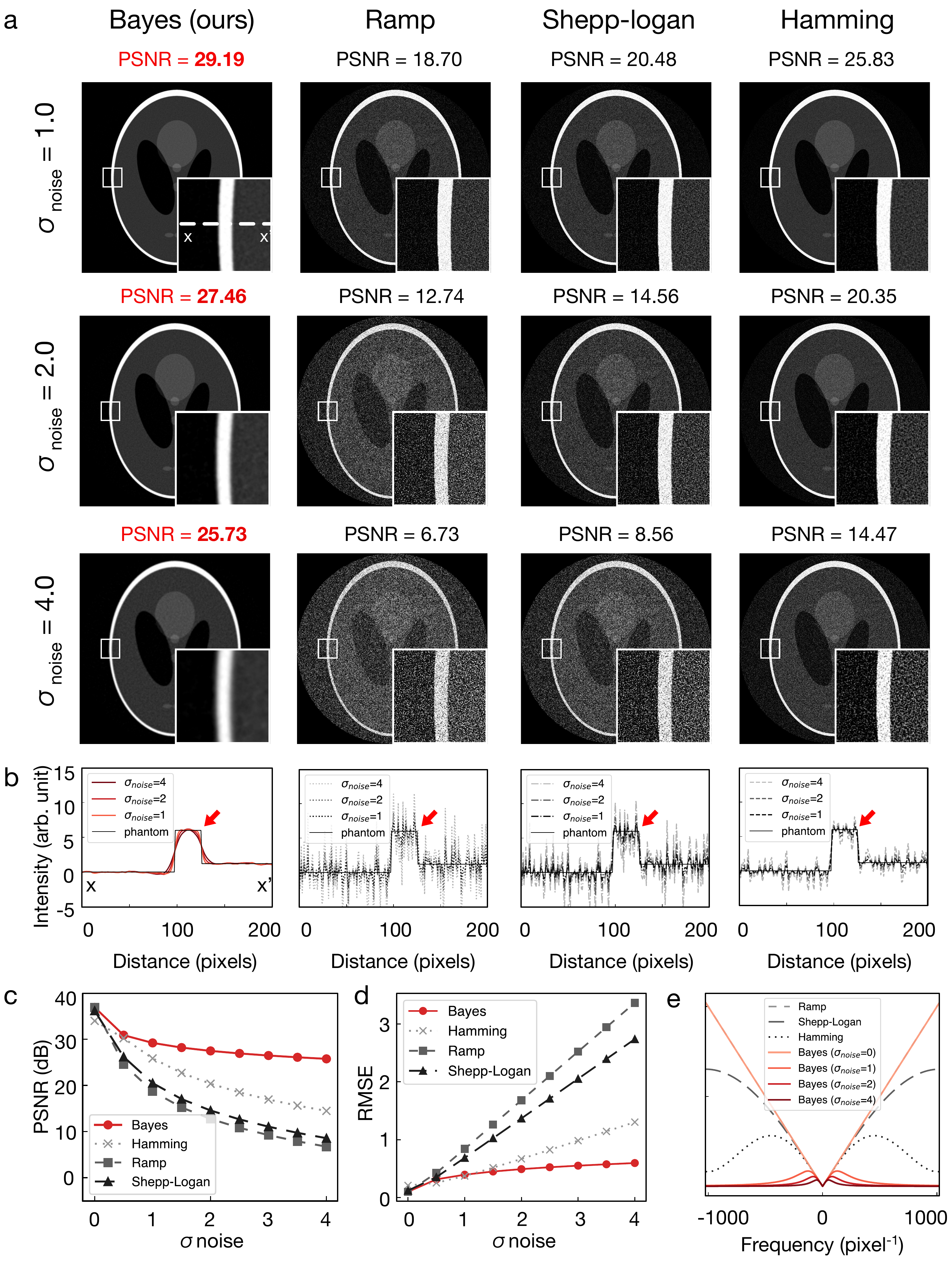}
  \caption{(Color online) Comparison of reconstruction results obtained by fixing the number of projections at $N_\theta = 1800$ and varying the noise intensity $\sigma_{\mathrm{noise}}$ as $1.0$, $2.0$, and $4.0$. (a) Reconstruction results of the proposed algorithm and FBP with different filters under each noise condition, together with magnified views of the regions indicated by the boxes. (b) Horizontal center line profiles of the magnified regions. In each panel, the profiles under the different noise conditions and the ground-truth profile of the phantom are shown as indicated in the legend. (c) PSNR of the reconstructed images as a function of the noise intensity. (d) RMSE of the center line profile in the magnified region. (e) Frequency characteristics of the reconstruction filters under the different noise conditions.}
  \label{res_3.2}
\end{figure}

We first evaluate the reconstruction performance by varying the noise intensity with the fixed number of projections ($N_\theta = 1800$).
\Cref{res_3.2}(a) shows the reconstructed images by the FBP and the proposed algorithm, together with magnified views of the regions indicated by the boxes. 
The reconstructed images show that the reconstruction performance of the FBP-based methods degrades as $\sigma_{\mathrm{noise}}$ increases.
Even at $\sigma_{\mathrm{noise}} = 1.0$, the reconstructions by the Ramp and the Shepp--Logan filters are degraded by granular noise.
In the high-noise condition ($\sigma_{\mathrm{noise}} = 4.0$), granular noise appears and the smooth structures are degraded in each FBP reconstruction.
This is because filters with fixed frequency responses used in FBP cannot sufficiently suppress high-frequency noise. The degradation is most pronounced in the reconstruction by the Ramp filter due to its enhancement of high-frequency components. 
In contrast, the proposed algorithm preserves the overall structure even at  $\sigma_{\mathrm{noise}} = 4.0$, demonstrating its robustness to noise. 

To further evaluate the reconstruction performance,  we examined the line profiles across the center of the magnified regions in \cref{res_3.2}(a).
As shown in \cref{res_3.2}(b), the profiles in the FBP reconstructions strongly oscillate as $\sigma_{\mathrm{noise}}$ increases, especially around the edge regions indicated by the arrows.
In contrast, the proposed algorithm suppresses the profile oscillations, and the overall intensity shows a good agreement with the ground-truth profile.
Filtering generally involves a trade-off between resolution and noise suppression, but the line profiles confirm that the noise-suppression effect in the proposed algorithm also works effectively from a local viewpoint.
Therefore, from both the full reconstructed image and the local line profiles, the proposed algorithm can be regarded as producing reconstruction results in which the influence of noise is suppressed more effectively than in FBP.

To quantify the reconstruction performance, we calculated PSNR for each reconstruction.
As shown in \cref{res_3.2}(a), at every $\sigma_{\mathrm{noise}}$, the proposed algorithm yields the highest PSNR; moreover, there is a clear difference in PSNR degradation as the noise intensity increases,
as shown in \cref{res_3.2}(c). Compared to FBP, the proposed algorithm exhibits only a small reduction in PSNR as the noise intensity increases, quantitatively demonstrating the high robustness to noise of the proposed algorithm.
The same tendency is confirmed in the reconstruction of local structures.
\Cref{res_3.2}(d) shows the RMSE between line profiles of reconstructed image and ground truth as a function of $\sigma_{\mathrm{noise}}$. The profile locations are the same as in \cref{res_3.2}(b).
In the reconstruction by FBP, RMSE increases as the noise intensity increases; in particular, the Ramp filter shows a rapid increase from $\sigma_{\mathrm{noise}} = 1.0$ to $4.0$.
In the reconstruction by the proposed algorithm, on the other hand, the increase in RMSE is much more moderate over the same interval, quantitatively demonstrating that it preserves local structures more accurately than FBP.

\Cref{res_3.2}(e) shows the filter shapes in the frequency domain.
At the small $\sigma_{\mathrm{noise}}$, the filter shape of the proposed algorithm is close to that of the Ramp filter, whereas stronger attenuation appears in the high-frequency region as the noise intensity increases.
This variation in the filter shape arises because the hyperparameters ($\gamma$, $\beta$, and $h$) estimated by free-energy minimization are varied according to the noise intensity.
In other words, the proposed algorithm adaptively changes its frequency response according to the noise intensity, which effectively reconstructs the overall image contrast while minimizing blurring of fine-scale features.

\subsubsection{Reconstruction performance under limited number of projections}
\begin{figure}[htbp]
  \centering
  \includegraphics[width=0.9\columnwidth]{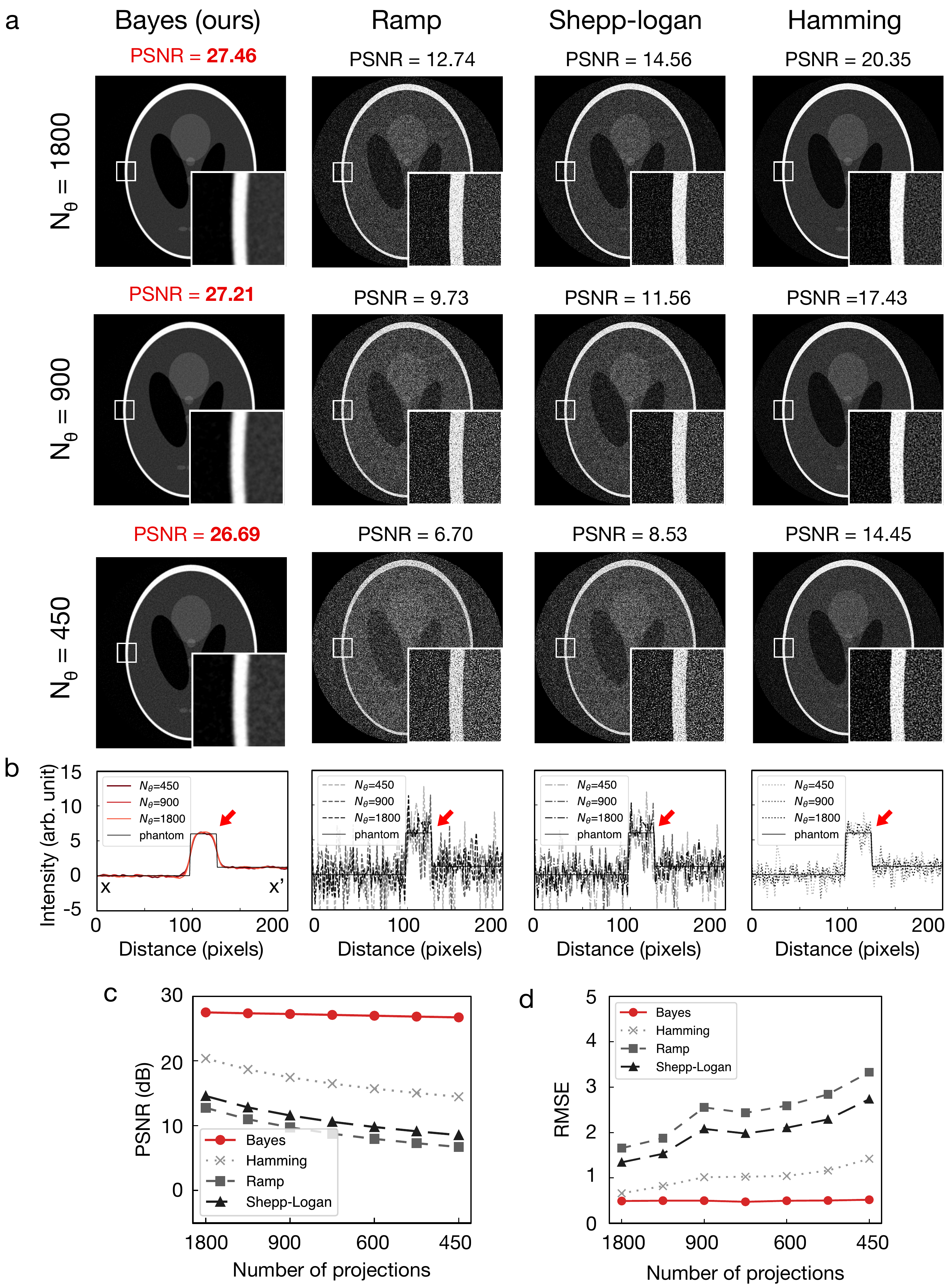}
    \caption{(Color online) Comparison of reconstruction results obtained by fixing the noise intensity at $\sigma_{\mathrm{noise}} = 2.0$ and varying the number of projections as $N_\theta = 1800$, $900$, and $450$. (a) Reconstruction results of the proposed algorithm and FBP with different filters for each projection number, together with magnified views of the regions indicated by the boxes. (b) Horizontal center line profiles of the magnified regions. In each panel, the profiles under the different projection conditions and the ground-truth profile of the phantom are shown as indicated in the legend. (c) PSNR of the reconstructed images as a function of the number of projections. (d) RMSE of the center line profile in the magnified region.}
  \label{res_3.3}
\end{figure}

Next, we evaluate the reconstruction performance by reducing the number of projections with the fixed noise intensity ($\sigma_{\mathrm{noise}}=2.0$).
\Cref{res_3.3}(a) shows the reconstructed images obtained by FBP and the proposed algorithm, together with magnified views of the regions indicated by the boxes.
The reconstructed images show that the reconstruction performance of the FBP-based reconstructions degrades as the number of projections decreases.
In particular, at $N_\theta = 450$, granular noise becomes prominent and smooth structures in flat regions are degraded.
As in Sect.~3.1, reducing the number of projections increases the effective noise component in the reconstructed image.
In contrast, even when $N_\theta$ is reduced to $450$, the proposed algorithm exhibits only a moderate increase in roughness, and structural boundaries remain largely preserved.

To further evaluate the reconstruction performance, we examined the line profiles across the center of the magnified regions in \cref{res_3.3}(a).
As shown in \cref{res_3.3}(b), the profiles in the FBP reconstructions oscillate as the number of projections decreases, especially around the edge regions indicated by the arrows.
In contrast, the proposed algorithm suppresses the profile oscillations, and the overall intensity shows a good agreement with the ground-truth profile.
Filtering generally involves a trade-off between resolution and noise suppression, but the line profiles confirm that the noise-suppression effect in the proposed algorithm also works effectively from a local viewpoint.
Therefore, from both the full reconstructed image and the local line profiles, the proposed algorithm can be regarded as yielding reconstructions that exhibit less degradation than FBP when the number of projections is reduced.

To quantify the reconstruction performance, we calculated PSNR for each reconstruction.
As shown in \cref{res_3.3}(a), at every projection setting, the proposed algorithm yields the highest PSNR; moreover, there is a clear difference in PSNR degradation when the number of projections is reduced,
as shown in \cref{res_3.3}(c).
For each of the three FBP filters, reducing $N_\theta$ from $1800$ to $450$ decreases PSNR by approximately $6\,\mathrm{dB}$.
For example, the Ramp filter decreases from $12.74\,\mathrm{dB}$ to $6.70\,\mathrm{dB}$, and the Shepp--Logan and Hamming filters show similar trends.
In contrast, the proposed algorithm decreases only from $27.46\,\mathrm{dB}$ to $26.69\,\mathrm{dB}$, corresponding to a reduction of about $0.8\,\mathrm{dB}$, which is much smaller than the approximately $6\,\mathrm{dB}$ decrease observed for FBP.
Compared to FBP, the proposed algorithm exhibits only a small reduction in PSNR as the number of projections is reduced, quantitatively demonstrating its robustness to sparse angular sampling.
The same tendency is confirmed in the reconstruction of local structures.
\Cref{res_3.3}(d) shows the RMSE between line profiles of reconstructed image and ground truth as a function of the number of projections.
The profile locations are the same as in \cref{res_3.3}(b).
In the reconstruction by FBP, RMSE increases as the number of projections decreases; for example, the Ramp filter increases by nearly a factor of two, from $1.384$ to $2.764$.
The Shepp--Logan and Hamming filters show similar increases.
In the reconstruction by the proposed algorithm, on the other hand, the increase in RMSE is much more moderate over the same interval, quantitatively demonstrating its higher robustness to a reduction in the number of projections than FBP.

It should be mentioned that angular undersampling actually introduces streak artifacts in the projection-reduction settings examined here.
Because the observation noise was fixed at $\sigma_{\mathrm{noise}} = 2.0$, the present experiments focused on reconstruction degradation under reduced-projection conditions with observation noise.
Within this condition, the proposed algorithm suppressed the degradation in both PSNR and local line profiles more effectively than FBP.

\section{Conclusion}
In this study, we formulated a CT image reconstruction method based on Bayesian inference by combining an observation model with an MRF-based prior, and estimated the hyperparameters by free-energy minimization.
Through numerical experiments using two ground-truth images, an MRF-generated image and a Shepp--Logan phantom, we demonstrated that the proposed algorithm achieves higher reconstruction performance than filtered back projection (FBP) under varying noise intensities and numbers of projections.
In particular, as the noise intensity increased, the proposed algorithm exhibited a much smaller decrease in PSNR than FBP for both images, indicating that it maintains high reconstruction performance even under high-noise conditions.
This improvement in the robustness to noise is attributed to the filter shape adaptively determined from the observed projected image through hyperparameter optimization.
Moreover, when the number of projections was reduced, the decrease in PSNR remained small for both images, unlike in FBP.
These results indicate that the proposed algorithm suppresses the degradation of reconstruction performance under both high-noise conditions corresponding to low dose and reduced-projection conditions, regardless of image type.
Future work includes improving CT reconstruction performance under even fewer projections by introducing different prior distributions and validating the proposed algorithm using experimentally measured CT data.

\begin{acknowledgment}
This work was supported by  JSPS KAKENHI Grant-in-Aid for Scientific Research (A), Japan (JP23H00486), Data Creation and Utilization-Type Material Research and Development Project, Japan by the Ministry of Education, Culture, Sports, Science, and Technology (MEXT) (JPMXP1122712807), and JST BOOST (JPMJBS2418).
\end{acknowledgment}

\appendix
\section{Mathematical Background of Filtered Back Projection}

Here, we review the Radon transform, the Fourier slice theorem, and filtered back projection (FBP), which form the mathematical basis of CT reconstruction \cite{Kak2001, Natterer2001}.

The Radon transform \cite{Radon1986} is defined as the line integral of $\xi(x,y)$ along the direction $t$ for each projection angle $\theta$ under the same coordinate system shown in \cref{theory}.
Here, $\xi(x,y)$ denotes the ground truth. The projected image $\tau_0(s,\theta)$ is given by
\begin{equation}\label{radon_def}
    \tau_0(s,\theta) = \int dt\,
    \xi(s\cos\theta - t\sin\theta,\, s\sin\theta + t\cos\theta).
\end{equation}
Thus, the reconstruction problem in CT is to recover $\xi(x,y)$ from the obtained projected image $\tau_0(s,\theta)$.
The inverse Radon transform is based on the Fourier slice theorem.
We first define the two-dimensional Fourier transform of $\xi(x,y)$ and its inverse transform as
\begin{equation}\label{ft2d}
    \tilde{\xi}(\tilde{x},\tilde{y}) = \iint dx\,dy\, \xi(x,y)\, e^{-2\pi i (\tilde{x}x + \tilde{y}y)},
\end{equation}
\begin{equation}\label{ift2d}
    \xi(x,y) = \iint d\tilde{x}\,d\tilde{y}\, \tilde{\xi}(\tilde{x},\tilde{y})\, e^{2\pi i (\tilde{x}x + \tilde{y}y)}.
\end{equation}
Here, $(\tilde{x},\tilde{y})$ denotes the spatial-frequency coordinates.
Similarly, the one-dimensional Fourier transform of the projected image $\tau_0(s,\theta)$ with respect to $s$ is defined as
\begin{equation}\label{ft1d}
    \tilde{\tau}_{0}(\tilde{s},\theta) = \int ds\, \tau_0(s,\theta)\, e^{-2\pi i \tilde{s} s}.
\end{equation}
By substituting \cref{radon_def} into \cref{ft1d} and transforming the integration variables from $(s,t)$ to $(x,y)$, we obtain
\begin{equation}\label{slice}
    \begin{split}
        \tilde{\tau}_{0}(\tilde{s},\theta)
        &= \int ds \int dt\,
        \xi(s\cos\theta - t\sin\theta,\, s\sin\theta + t\cos\theta)\, e^{-2\pi i \tilde{s} s} \\
        &= \iint dx\,dy\, \xi(x,y)\, e^{-2\pi i \tilde{s}(x\cos\theta + y\sin\theta)} \\
        &= \tilde{\xi}(\tilde{s}\cos\theta,\, \tilde{s}\sin\theta).
    \end{split}
\end{equation}
Therefore, the one-dimensional Fourier transform of the projected image at angle $\theta$ is equal to the two-dimensional Fourier transform of $\xi(x,y)$ on the radial line passing through the origin at angle $\theta$.
If the projected image is obtained over $\theta \in [0,\pi)$, the Fourier components $\tilde{\xi}(\tilde{x},\tilde{y})$ can be determined from $\tilde{\tau}_{0}(\tilde{s},\theta)$.
This relation is called the Fourier slice theorem.

FBP is derived by rewriting the two-dimensional inverse Fourier transform (\cref{ift2d}) in polar coordinates.
By changing variables as $(\tilde{x},\tilde{y}) = (\tilde{s}\cos\theta,\, \tilde{s}\sin\theta)$, we obtain $d\tilde{x}\,d\tilde{y} = |\tilde{s}|\,d\tilde{s}\,d\theta$; thus,
\begin{equation}\label{fbp}
    \xi(x,y) = \int_0^{\pi} d\theta \int_{-\infty}^{\infty} d\tilde{s}\,
    |\tilde{s}|\, \tilde{\xi}(\tilde{s}\cos\theta,\, \tilde{s}\sin\theta)\,
    e^{2\pi i \tilde{s}(x\cos\theta + y\sin\theta)}.
\end{equation}
Using the Fourier slice theorem (\cref{slice}), $\tilde{\xi}(\tilde{s}\cos\theta,\, \tilde{s}\sin\theta)$ can be replaced by the Fourier-transformed projected image $\tilde{\tau}_{0}(\tilde{s},\theta)$:
\begin{equation}\label{fbp2}
    \xi(x,y) = \int_0^{\pi} d\theta \int_{-\infty}^{\infty} d\tilde{s}\,
    |\tilde{s}|\, \tilde{\tau}_{0}(\tilde{s},\theta)\,
    e^{2\pi i \tilde{s}(x\cos\theta + y\sin\theta)}.
\end{equation}
By defining $g(s,\theta)$ as
\begin{equation}\label{filtered_proj}
    g(s,\theta) = \int d\tilde{s}\,
    |\tilde{s}|\, \tilde{\tau}_{0}(\tilde{s},\theta)\, e^{2\pi i \tilde{s} s},
\end{equation}
\cref{fbp2} can be given by
\begin{equation}\label{backproj}
    \xi(x,y) = \int_0^{\pi} d\theta\, g(x\cos\theta + y\sin\theta,\, \theta),
\end{equation}
which indicates that the ground truth is reconstructed by applying the frequency-domain filter $|\tilde{s}|$ to the projected image and performing back projection.
This reconstruction procedure is called FBP.

Using the convolution theorem, \cref{filtered_proj} can also be expressed as a convolution in real space:
\begin{equation}\label{conv}
    g(s,\theta) = \int du\, h(u)\, \tau_0(s-u,\theta),
\end{equation}
where
\begin{equation}\label{ramp_kernel}
    h(u)
    =
    \int d\tilde{s}\,
    |\tilde{s}|\,e^{2\pi i\tilde{s}u}.
\end{equation}
Here, $h(u)$ is the real-space impulse response corresponding to
the Ramp filter $|\tilde{s}|$ in the frequency domain.
Therefore, in FBP, the design of the filter function $h(u)$, or equivalently its frequency response, is an important factor to determine the reconstruction performance.
Ramachandran and Lakshminarayanan proposed the Ramp filter \cite{Ramachandran1971}, and Shepp and Logan further proposed a modified filter that suppresses high-frequency components \cite{Shepp1974}.

\begin{figure}[htbp]
  \centering
  \includegraphics[width=0.45\columnwidth]{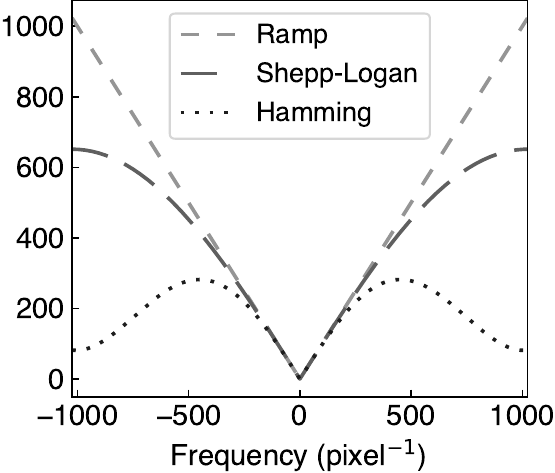}
  \caption{Frequency responses of the Ramp, Shepp--Logan, and Hamming filters used in FBP.}
  \label{fbp_filter}
\end{figure}

In practice, observation noise is superimposed on the projected image, as modeled in the main text.
As shown in \cref{fbp_filter}, since the Ramp filter increases linearly with frequency, it amplifies the high-frequency components of the projected image, which could cause granular noise in the reconstructed image.
This degradation becomes particularly pronounced under low-dose or sparse-view conditions.
The Shepp--Logan and Hamming filters \cite{Shepp1974, Harris1978} mitigate the degradation by multiplying the Ramp filter with window functions that attenuate high-frequency components, as shown in \cref{fbp_filter}.
However, since their frequency response is fixed regardless of the noise intensity and experimental conditions, they are not guaranteed to be optimal filter under all observation conditions.
In contrast, this limitation should be solved in the proposed method, in which the filter shape is adaptively determined from the observed projected image and it is optimized for each experimental condition.

\end{document}